\documentstyle[12pt,epsf]{article}
\newcommand{\sss}{\scriptscriptstyle}
\newcommand{\pl}{Phys.\ Lett.} 

\newcommand{\prl}{Phys.\ Rev.\ Lett.} 
\newcommand{\prd}{Phys.\ Rev.\ D}

\newcommand{\etal}{{\it et al}.,\ }
\newcommand{\real}{{\rm Re}\,}
\newcommand{\imag}{{\rm Im}\,}
\def\be{\begin{equation}}
\def\ee{\end{equation}}
\def\bea{\begin{eqnarray}}
\def\eea{\end{eqnarray}}
\def\fH{{\cal H}}

\begin{document}

\pagestyle{empty}

\vspace{4.truecm} 
\begin{center} {\large  \bf FLAVOR CHANGING NEUTRAL CURRENTS
AND THE THIRD FAMILY}\end{center}
\vspace{1.truecm}
\begin{center}
Laura Reina \\ 
\vspace{.5truecm}
Physics Department, Brookhaven National Laboratory, Upton, NY\ \ 11973
\end{center} 

\vspace{5.truecm}

\begin{quote}{\bf Abstract}: We consider a Two Higgs Doublet Model with Flavor
Changing Scalar Neutral Currents arising at the tree level. All the
most important constraints are taken into account and the
compatibility with the present Electroweak measurements is
examined. The Flavor Changing couplings involving the third family are
not constrained to be very small and this allows us to predict some
interesting signals of new physics. (This paper relies on some work
done in collaboration with D. Atwood (CEBAF) and A. Soni (BNL)). 
\end{quote}
\vspace{1truein}
\begin{center}
{\it to appear in the Proceedings of the XXXIst Rencontres de Moriond},
{\it ``Electoweak Interactions'', Les Arcs, France, March 1996}.
\end{center}

\newpage
\setcounter{page}{1}
\pagestyle{plain}

All processes involving Flavor Changing Neutral Currents (FCNC) are
suppressed in the Standard Model (SM) because they are forbidden at
the tree level. Some of them end up having a measurable, although
small, branching fraction since they are enhanced at the loop level by
the presence of a top quark in the loop. This is the case of some
radiative $B$-meson decays, like those induced at the parton level by
$b\rightarrow s\gamma$ ($Br(B\rightarrow X_s\gamma)\sim 10^{-4}$
\cite{alam}). However, a similar enhancement cannot take place for the
up-type FC transitions and therefore this can be a good place to look
for evidence of new physics.

Moreover, the outstanding nature of the top quark (with its huge mass,
$m_t\sim 175$ GeV) should induce us to reexamine our theoretical
prejudices about the existence of Flavor Changing Scalar Interactions
(FCSI), expecially for the top quark itself. Probing the top-charm and
top-up flavor changing vertex consequently deserves a special
attention. We will present a theoretical model in which FCSI can be
generated at the tree level with a given hierarchy and discuss some
possible experimental environments in which definite bounds on the top
quark FC couplings can be put.

We will consider a Two Higgs Doublet Model (2HDM) with allowed FCNC in
the scalar sector, the so called Model III \cite{savage}. In fact, in
models with a non-minimal Higgs sector, e.g. in the 2HDM, FCSI arise
readily at the tree level. In order to avoid the severe constraints
from $K^0\!-\!\bar K^0$ and $B^0\!-\!\bar B^0$ mixing, it was
originally proposed \cite{glash} to forbid all FCSI by imposing a
suitable discrete symmetry acting on the quark and the scalar fields
\cite{kraw}.  However, as later realized by many authors
\cite{sheretal}, it is possible to remove the {\it ad hoc} discrete
symmetry and satisfy the constraints by chosing an adequate ansatz on
the FC couplings. In particular, it was observed that the necessary
hierarchy on the FC couplings between fermions and scalars is provided
by the mass parameters of the fermion fields themselves
\cite{sheretal}. If this is the case, then the top quark FC couplings
can be greatly enhanced with respect to the first and second
generation ones.

In some recent papers \cite{eetc,mumutc,rbrc}, we have analyzed in
detail this kind of 2HDM and studied the possible phenomenological
implications that large FC top couplings can have. Due to the
theoretical and experimental interest of this analysis, we want to
provide a brief but comprehensive description of Model III and of the
most important constraints that affect its FCSI. Given the constrained
Model, we will proceed to the discussion of some clean experimental
environments in which signals from FCSI can be detected.

\vspace{.5truecm} Let us focus on the quark Yukawa interactions
only and write the corresponding Yukawa Lagrangian for a 2HDM in the
following very general form \cite{savage}

\be {\cal L}_{Y} =  \eta^{\sss U}_{ij} \bar Q_i \tilde\phi_1 U_j + 
\eta^{\sss D}_{ij} \bar Q_i\phi_1 D_j + 
\xi^{\sss U}_{ij} \bar Q_i\tilde\phi_2 U_j
+\xi^{\sss D}_{ij}\bar Q_i \phi_2 D_j \,+\, h.c.
\label{lyuk}
\ee

\noindent where $\phi_i$ for $i=1,2$ are the two scalar doublets,
while $\eta^{\sss U,D}_{ij}$ and $\xi_{ij}^{\sss U,D}$ are the non diagonal
coupling matrices. By a suitable rotation of the fields we chose the
physical scalars in such a way that only the $\eta_{ij}^{\sss U,D}$
couplings generate the fermion masses, i.e. such that

\be
<\phi_1>=\left(
\begin{array}{c}
0\\
{v/\sqrt{2}}
\end{array}
\right), \ \ \ \ 
<\phi_2>=0 
\ee 

\noindent The physical spectrum consists of two charged
$\phi^{\pm}$ and three neutral spin 0 bosons, two scalars ($H^0,h^0$)
and a pseudoscalar ($A^0$)

\bea 
\label{neut}
H^0 & = & \sqrt{2}[({\real}\phi^0_1-v) \cos \alpha+ \real\phi^0_2 \sin \alpha ]
\nonumber \\ 
h^0 & = & \sqrt{2}[-(\real\phi^0_1-v) \sin\alpha +\real\phi^0_2
\cos\alpha ] \\ 
A^0 & = & \sqrt{2} (\imag\phi^0_2) \nonumber 
\eea 

\noindent where $\alpha$ is a mixing phase (for $\alpha\!=\!0$, $H^0$
corresponds exactly to the SM Higgs field, and $\phi^{\pm}$, $h^0$ and
$A^0$ generate the new FC couplings). In principle the
$\xi_{ij}^{\sss U,D}$ FC couplings are arbitrary, but reasonable arguments
exist to adopt the following ansatz \cite{sheretal}

\be 
\xi^{\sss U,D}_{ij} = \lambda_{ij}\frac{\sqrt{m_im_j}}{v}
\label{csi}
\ee 

\noindent where for the sake of simplicity we take the $\lambda_{ij}$
parameters to be real (for more details see \cite{eetc,rbrc}).
Alternatively, we can assume the $\xi^{\sss U,D}_{ij}$ to be purely
phenomenological couplings and try to constrain them from experiments.
In fact, the two assumptions are almost equivalent, as far as one keeps
a certain arbitrarity on the parameters $\lambda_{ij}$ of
eq.(\ref{csi}).  What is really crucial to our analysis is to derive a
consistent scenario in which each FC coupling of Model III is
constrained from some existing phenomenology.

{}From a detailed analysis\footnote{In our analysis
the masses of the scalar particles are let vary in the range $200$
GeV$<M_s<1$ TeV and the phase $\alpha$ is taken to be
$\alpha=0,\pi/4$.}  \cite{eetc}, we obtain that $\xi_{ds}$ and
$\xi_{db}$ are constrained to be very small by the experimental
measurement of the $K^0\!-\!\bar K^0$ ($\Delta M_K^{exp}=3.51\times
10^{-15}\,\mbox{GeV}$) and $B^0\!-\!\bar B^0$ ($\Delta
M_B^{exp}=3.36\times 10^{-13}\,\mbox{GeV}$) mixings\footnote{In a
Model with FCSI a process like $F^0\!-\!\bar F^0$ mixing occurs at the
tree level and, unless unusual assumptions on the FC couplings are
made, this constitutes the leading contribution.}. To a less extent
the $D^0\!-\!\bar D^0$ mixing ($\Delta M_D^{exp}\le 1.32\times
10^{-13}\,\mbox{GeV}$) also constrains $\xi_{uc}$ to be small and it
is likely to give a bound as severe as the ones from the previous two
mixings as soon as the experimental precision improves by an extra
order of magnitude.  Almost all the FC couplings involving the first
generation are therefore immediately suppressed, confirming the
hierarchical nature of the FCSI of Model III. Hence, we can make the
more general assumption that all the FC couplings involving the first
generation are almost negligible, extending the previous bounds also
to $\xi^{\sss U,D}_{ut}$, even if we do not have any direct constraint
at the moment.

The loop contributions to the previous $F^0\!-\!\bar F^0$ mixings and
the $Br(B\rightarrow X_s\gamma)$ put limits on the remaining
$\xi^{\sss U,D}_{ij}$ couplings. In particular, it turns out that they
can be well described by the ansatz in eq.(\ref{csi}), when the
corresponding $\lambda_{ij}$ parameters vary in the range
$0.1<\lambda_{ij}<10$. In this way the hierarchy of the FC couplings is
still garantied. 

Moreover, the analysis of the $Br(B\rightarrow X_s\gamma)$ and of the
$\rho$ parameter\footnote{We recall that
$\rho=M_W^2/(c_W^2M_Z^2)(1+\Delta\rho_0)$ where $\Delta\rho_0$
parametrize the deviation from the SM result.} (see ref. \cite{rbrc}
for full details) selects a specific region of the scalar mass
parameter space.  The charged scalar mass has to be $M_c>600$ GeV and
one of the following conditions

\be
M_H,M_h\le M_c\le M_A\,\,\,\,\,\mbox{and}\,\,\,\,
M_A\le M_c\le M_H,M_h
\label{Mc_inbetween}
\ee

\noindent has to be verified, where $M_H$ and $M_h$ are the masses of
the neutral scalars $H^0$ and $h^0$ and $M_A$ is the mass of the
neutral pseudoscalar $A^0$, as in eq.(\ref{neut}).

\vspace{.5truecm} Within this version of Model III we can
now draw the attention to some interesting process that could help
to constrain the third family FC couplings. We think that top-charm
production at a high energy lepton collider, i.e. $e^+e^-,\,
\mu^+\mu^-\rightarrow \bar t c+\bar c t$, can be very distinctive
under many respects. First of all, this is the kind of process whose
SM prediction is extremely suppressed \cite{eetc} and any signal would
be a clear evidence of new physics with large FC couplings in the
third family. Second, it has a very clean kinematical signature, with
a very massive jet recoiling against an almost massless one (very
different from a $bs$ FC signal, for instance). This characteristic is
enhanced even more in the experimental environment of a lepton
collider, because of the very low background. A part from these
general considerations, the cases of an $e^+e^-$ collider and of a
$\mu^+\mu^-$ collider require a separate analysis.

At an $e^+e^-$ collider the $tc$-production process arises at the one
loop level via $e^+e^-\rightarrow\gamma^*,Z^*\rightarrow\bar tc+\bar
ct$. The tree level FC processes generated by the s-channel exchange
of a scalar boson ($h^0,\ldots$) are suppressed due to the smallness of
the electron mass. The effective $\gamma tc$ and $Ztc$ vertices can be
calculated at one loop \cite{eetc} and used in the calculation of the
cross section. Of particular relevance is the normalized ratio

\be 
R_{tc}\equiv\frac{\sigma(e^+e^-\rightarrow t\bar c+ \bar tc)} 
{\sigma( e^+e^- \rightarrow\gamma\rightarrow\mu^+\mu^-)} 
\label{rtc_ee}
\ee 

\begin{figure}[htb]
\centering
\epsfxsize=5.in
\leavevmode\epsffile{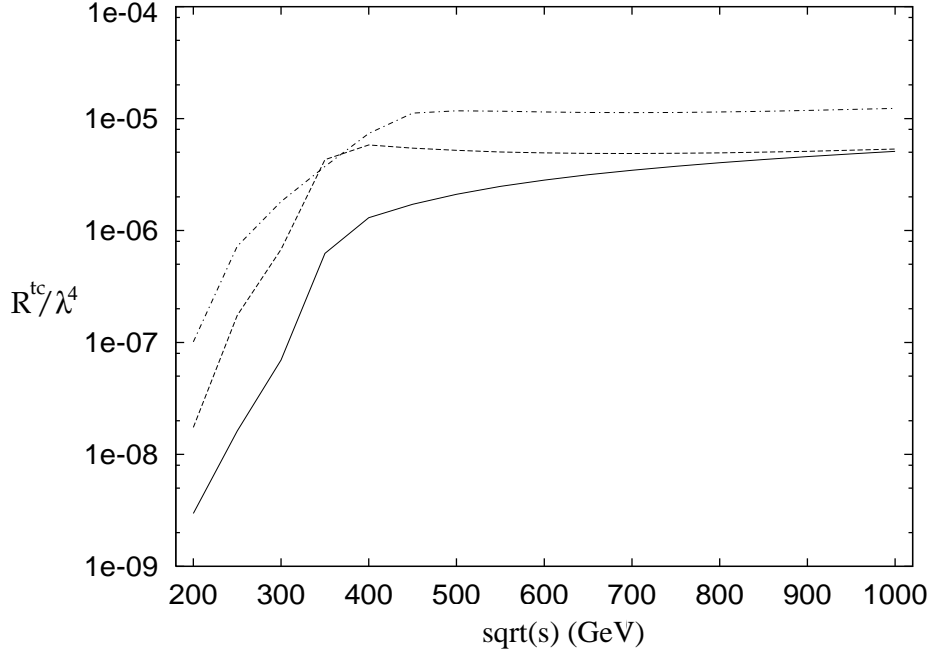}
\caption[]{ $R_{tc}/\lambda^4$ vs.\ $\sqrt{s}$ when $M_h\!=\!200$ GeV
and $M_A\!\simeq\!M_c\!=\!1$ TeV (solid), $M_A\!=\!200$ GeV and
$M_h\!\simeq\!M_c\!=\!1$ TeV (dashed), $M_c\!=\!200$ GeV and
$M_h\!\simeq\!M_A\!=\!1$ TeV (dot-dashed).}
\label{eetc}
\end{figure}

\noindent We will assume for purpose of illustration a common value of
$\lambda_{ij}$ for all the $\xi^{\sss U,D}_{ij}$ couplings
involved. The ratio $R_{tc}$ scales as $\lambda^4$, therefore our
predictions crucially depend on the value of the arbitrary parameter
$\lambda$. In particular, $R_{tc}$ is governed, in the mass parameter
range of eq.(\ref{Mc_inbetween}), by $\xi_{tt}$ and $\xi_{tc}$ and
$e^+e^-\rightarrow \bar tc+\bar ct$ will be an important process to
constrain the magnitude of these non standard couplings.  As an
example we plot in Fig. \ref{eetc} the ratio $R_{tc}$ normalized to
$\lambda^4$ as a function of $\sqrt{s}$. We look in particular to the
case in which one scalar mass is light ($M_l\sim 200$ GeV) and the
other two are much heavier ($M_h\sim 1$ TeV), which could correspond
to one of the two conditions in eq.(\ref{Mc_inbetween}). As we can see
from Fig. \ref{eetc} there is no need to increase $\sqrt{s}$ above 500
GeV, where $10^4-10^5\,\mu^+\mu^-$ are predicted. We can see that for
values of $\lambda$ a little bigger than unity a few events can be
seen. Since there is no experimental basis for assuming the absence of
tree level FCNC at the scale $m_t$, their rigorous search is strongly
advocated.

Another interesting possibility to study top-charm production is
offered by Muon Colliders \cite{mumutc}. Although very much in the
notion stage at present, $\mu^+\mu^-$ colliders has been suggested as
a possible lepton collider for energies in the TeV range
\cite{palmer}.  Most of the applications of Muon Colliders would be
very similar to electron colliders. One advantage, however, is that
they may be able to produce Higgs bosons ($\fH$) in the $s$ channel in
sufficient quantity to study their properties directly (remember that
$m_{\mu}\simeq 200\, m_e$). The crucial point is also that in spite of
the fact that the $\mu^+\mu^-\fH$ coupling, being proportional to
$m_\mu$, is still small, if the Muon Collider is run on the Higgs
resonance, $\sqrt{s}=m_\fH$, Higgs bosons may be produced at an
appreciable rate.

We have considered \cite{mumutc} the simple but fascinating
possibility that such a Higgs, $\fH$, has a flavor-changing $\fH t\bar
c$ coupling, as is the case in Model III or in any other 2HDM with
FCNC. As we did for the $e^+e^-$ case, also in the $\mu^+\mu^-$ case
we can define the analogous of $R_{tc}$ in eq.(\ref{rtc_ee}) to be

\begin{equation}
\tilde R_{tc}= \tilde R(\fH)\,(B^\fH_{t\bar c}+B^\fH_{c\bar t})
\label{rtc_mumu}
\end{equation}

\noindent where $\tilde R(\fH)$ is the effective rate of Higgs
production at a Muon Collider and $B^\fH_{t\bar c}$ or $B^\fH_{c\bar
t}$ denote the branching ratio for $\fH\rightarrow t\bar c$ and
$\fH\rightarrow c\bar t$ respectively.  Assuming that the background
will be under reasonable control by the time they will start operate a
Muon Collider, our extimate is that $10^{-3}<\tilde R_{tc}\le 1$,
depending on possible different choices of the parameters. For a Higgs
particle of $m_\fH=300$ GeV, a luminosity of $10^{34} cm^{-2} s^{-1}$
and a year of $10^7 s$ ($1/3$ efficiency), a sample of $tc$ events
ranging from almost one hundred to few thousands can be produced
\cite{mumutc}. Given the distinctive nature of the final state and the
lack of a Standard Model background, the predicted luminosity should
allow the observation of such events. Therefore many properties of the
Higgs-tc coupling could be studied in detail.

\end{document}